\documentstyle[seceq,preprint]{ptptex}

\newcommand{\bra}[1]{\langle {#1} |}     
\newcommand{\ket}[1]{| {#1} \rangle}     
\newcommand{\rket}[1]{| {#1} )}     
\newcommand{\dket}[1]{|\!| {#1} \rangle}     

\title{
Deformed Boson Scheme Stressing Even-Odd \\
Boson Number Difference. II
}
\subtitle{Unified Forms of Boson-Pair Coherent States in Even- and 
Odd-Boson Systems
}    

\author{
Atsushi {\sc Kuriyama},$^{1}$ 
Constan\c{c}a {\sc Provid\^encia},$^{2}$ \\
Jo\~ao da {\sc Provid\^encia},$^{2}$ Yasuhiko {\sc Tsue}$^{3}$ 
and Masatoshi {\sc Yamamura}$^{1}$
}

\inst{
$^1$Faculty of Engineering, Kansai University, Suita 564-8680, Japan\\
$^{2}$Departamento de Fisica, Universidade de Coimbra, P-3000 Coimbra, 
Portugal\\
$^{3}$Physics Division, Faculty of Science, Kochi University, Kochi 780-8520, 
Japan
}

\recdate{
\today
}

\abst{
The boson-pair coherent state developed in Part (I) is generalized 
to the case in which the state is a mixture of even-boson and 
odd-boson number states. 
A general framework is shown and its three concrete examples are discussed. 
Main idea is in the application of the deformed boson scheme 
presented by the present authors. 
}

\begin{document}

\maketitle

\section{Introduction}

In Part (I),\cite{one} various forms of boson-pair coherent states were 
investigated. The main aim of this study 
is to describe the time-evolution of many-boson 
system in terms of the time-dependent variational method. 
The boson-pair coherent states play the role of the trial states for the 
variation. In (I), we treated two types of the boson-pair 
coherent states separately, but intimately. 
One type consists of the orthogonal states with even-boson number 
and the other with odd-boson number. 
With the aid of this treatment, we can describe the time-evolution of 
many even- and odd-boson states separately, but intimately. 
However, the conventional boson coherent state consists of the 
orthogonal states with the even- and the odd-boson numbers, and 
then, it may be powerful for describing the case in which the 
initial state is a mixture of the even- and the odd-boson numbers. 
But, the mixture is restricted to be of a special form.

The main aim of the present paper, Part (II), is to present a 
possible form of the boson coherent state which consists of the 
orthogonal states with the even- and the odd-boson states. 
This is a superposition of the boson-pair states $\ket{ch}$ and 
$\ket{sh}$. The coefficients of the superposition are given 
by new parameters and their special case corresponds to the 
conventional boson coherent state. 
The basic idea comes from that proposed by the present authors 
in Ref.\citen{two}. 
In Ref.\citen{two}, we investigated the boson-pair 
coherent state under a close connection to the $su(1,1)$-spin with the 
magnitudes $t=1/4$ and $3/4$. 
In the present paper, this case is generalized to any one. 
The idea is based on the deformed boson scheme investigated by the 
present authors.\cite{three} 
After a general framework is formulated, three examples which are 
treated in detail in Part (I) are investigated in concrete forms. 
In these treatments, Refs.\citen{four} and \citen{five} play 
an essential role.

In \S 2, a general framework is presented. Section 3 is devoted to giving 
two examples which lead to an exact solution and an approximated 
one with high accuracy for the basic equations given in \S 2. 
One is closely related to the $su(1,1)$-algebra. In \S 4, a straightforward 
generalization from the conventional boson coherent state 
is discussed in a possible approximated form. 
Further, in \S 5, the forms discussed in \S\S 2, 3 and 4 are 
formulated in terms of the MYT boson mapping. 
Finally, in \S 6, the case discussed in \S 4 is reinvestigated 
in a form simpler than that given in \S 4.

\section{General framework}

With the recapitulation of Part (I), let us, first, give a basic 
framework of our present form. 
In (I), we investigated two states $\ket{ch}$ and $\ket{sh}$ 
consisting of one kind of boson operator $({\hat c}, {\hat c}^*)$ 
separately, but intimately. They are defined as follows : 
\begin{subequations}\label{2-1}
\begin{equation}
\ket{ch}=\left(\sqrt{\Gamma_{ch}}\right)^{-1}\dket{ch} \ , \quad\qquad
(\bra{ch}ch\rangle =1) \qquad\quad
\label{2-1a}
\end{equation}
\end{subequations}
\vspace{-0.2cm}
\begin{subequations}\label{2-2}
\begin{equation}
\dket{ch}=\ket{0}+\sum_{n=1}^{\infty}\frac{\gamma^{2n}}{\sqrt{(2n)!}}
{\tilde f}(0){\tilde f}(1)\cdots {\tilde f}(2n-1)\ket{2n} 
\ , 
\label{2-2a}
\end{equation}
\end{subequations}
\vspace{-0.2cm}
\begin{subequations}\label{2-3}
\begin{equation}
\Gamma_{ch}=1+\sum_{n=1}^{\infty}\frac{(|\gamma|^{2})^{2n}}{{(2n)!}}
\left({\tilde f}(0){\tilde f}(1)\cdots {\tilde f}(2n-1)\right)^2
\ , 
\label{2-3a}
\end{equation}
\end{subequations}
\setcounter{equation}{0}
\begin{subequations}
\begin{equation}
\ket{sh}=\left(\sqrt{\Gamma_{sh}}\right)^{-1}\dket{sh} \ , \quad\qquad
(\bra{sh}sh\rangle =1) \qquad\quad
\label{2-1b}
\end{equation}
\end{subequations}
\vspace{-0.2cm}
\begin{subequations}
\begin{equation}
\dket{sh}=\sum_{n=0}^{\infty}\frac{\gamma^{2n+1}}{\sqrt{(2n+1)!}}
{\tilde f}(0){\tilde f}(1)\cdots {\tilde f}(2n)\ket{2n+1} 
\ , 
\label{2-2b}
\end{equation}
\end{subequations}
\vspace{-0.2cm}
\begin{subequations}
\begin{equation}
\Gamma_{sh}=\sum_{n=0}^{\infty}\frac{(|\gamma|^{2})^{2n+1}}{{(2n+1)!}}
\left({\tilde f}(0){\tilde f}(1)\cdots {\tilde f}(2n)\right)^2
\ . \qquad
\label{2-3b}
\end{equation}
\end{subequations}
Here, $(\gamma, \gamma^*)$ denotes a set of complex parameters and 
$\ket{k}$ ($k=2n,\ 2n+1$) is defined as 
\begin{equation}\label{2-4}
\ket{k}=(1/\sqrt{k!}) ({\hat c}^*)^k \ket{0} \ , \qquad 
({\hat c}\ket{0}=0) \ .
\end{equation}
The function ${\tilde f}(k)$ characterizes a deformation of boson operator 
and obeys 
\begin{equation}\label{2-5}
{\tilde f}(0)=1, \qquad {\tilde f}(k) > 0 \ , \qquad (k=1, 2, 3, \cdots) \ .
\end{equation}
It may be self-evident that the states $\ket{ch}$ and $\ket{sh}$ 
consist of even- and odd-boson number states, respectively.

Main aim of this paper, Part (II), is to investigate 
a unified form of $\ket{ch}$ and $\ket{sh}$. First, we introduce 
a state $\ket{cr}$ defined as 
\begin{equation}\label{2-6}
\ket{cr}=u\ket{ch}+(\gamma^*/|\gamma|) v \ket{sh} \ , 
\qquad (\bra{cr}cr\rangle =1) 
\end{equation}
Here, $(v, v^*)$ denotes a set of complex parameters additional to the 
set $(\gamma, \gamma^*$). The quantity $u$ is real and obeys 
\begin{equation}\label{2-7}
u^2+|v|^2=1 \ . 
\end{equation}
If $v=0$ or $u=0$ $(v=\gamma/|\gamma|)$, $\ket{cr}$ is reduced to 
$\ket{ch}$ or $\ket{sh}$, respectively. 
Further, $\ket{cr}$ is reduced to the state $\ket{ex}$ introduced in the 
relation (4.1) in (I) under the condition 
\begin{equation}\label{2-8}
u=\sqrt{\Gamma_{ch}/(\Gamma_{ch}+\Gamma_{sh})} \ , \qquad
v=(\gamma/|\gamma|)\sqrt{\Gamma_{sh}/(\Gamma_{ch}+\Gamma_{sh})} \ .
\end{equation}
The above tells us that $\ket{cr}$ at three limits is reduced 
to the three types of the states which we have already investigated. 
In this sense, the state $\ket{cr}$ defined in the relation (\ref{2-6}) 
is a possible unified form of $\ket{ch}$ and $\ket{sh}$. 
It should be noted that the state $\ket{cr}$ is reduced to 
the conventional boson coherent state if ${\tilde f}(k)=1$ under the condition 
(\ref{2-8}). 
The above permits us to conclude that $\ket{cr}$ is a state stressing 
even-odd boson number difference.

After rather tedious calculation, the following 
relation is derived : 
\begin{eqnarray}\label{2-9}
\bra{cr}\partial_z\ket{cr}&=&(1/2)\cdot (v^*\cdot \partial_z v-v\cdot \partial_z v^*) \nonumber\\
& &+(1/2)\cdot (\gamma^*\cdot \partial_z \gamma 
-\gamma\cdot \partial_z \gamma^*) \nonumber\\
& &\times \left[ u^2(\Gamma'/\Gamma)_{ch}+|v|^2\left(
(\Gamma'/\Gamma)_{sh}-|\gamma|^{-2}\right)\right] \ ,
\end{eqnarray}
\vspace{-0.2cm}
\setcounter{equation}{8}
\begin{subequations}
\begin{equation}
(\Gamma'/\Gamma)_{ch}
=\frac{d\Gamma_{ch}}{d|\gamma|^2}\cdot \Gamma_{ch}^{-1} \ , \qquad
(\Gamma'/\Gamma)_{sh}
=\frac{d\Gamma_{sh}}{d|\gamma|^2}\cdot \Gamma_{sh}^{-1} \ . 
\label{2-9a}
\end{equation}
\end{subequations}
Instead of two sets $(\gamma, \gamma^*)$ and $(v, v^*)$, 
we introduce two sets $(c,c^*)$ and ($\eta, \eta^*)$ 
which obey 
\begin{eqnarray}
& &\bra{cr}\partial_c \ket{cr}=(1/2)c^* \ , \qquad
\bra{cr}\partial_{c^*} \ket{cr}=-(1/2)c \ , 
\label{2-10}\\
& &\bra{cr}\partial_\eta \ket{cr}=(1/2)\eta^* \ , \qquad
\bra{cr}\partial_{\eta^*} \ket{cr}=-(1/2)\eta \ . 
\label{2-11}
\end{eqnarray}
The parameters $(c, c^*)$ and $(\eta, \eta^*)$ obeying the conditions 
(\ref{2-10}) and (\ref{2-11}) can be regarded as canonical 
variables in classical mechanics.\cite{four} 
Our present problem is to express 
$(\gamma, \gamma^*)$ and $(v, v^*)$ in terms of $(c, c^*)$ and 
$(\eta, \eta^*)$. 
For this purpose, we set up the following forms for 
$\gamma$ and $v$ : 
\begin{eqnarray}
& &\gamma=\sqrt{\sqrt{2}c}\cdot\sqrt[4]{F_{cr}} \ , 
\label{2-12}\\
& &v=\eta\cdot G_{cr} \ . 
\label{2-13}
\end{eqnarray}
Here, $F_{cr}$ and $G_{cr}$ are functions of $|c|^2$ and 
$|\eta|^2$ (or $|\gamma|^2$ and $|v|^2$), the explicit forms of which 
should be determined. 
The forms (\ref{2-12}) and (\ref{2-13}) give us 
\begin{eqnarray}\label{2-14}
& &\gamma^*\cdot \partial_c \gamma - \gamma\cdot \partial_{c}\gamma^*
=c^*\cdot(1/\sqrt{2|c|^2})\sqrt{F_{cr}} \ , 
\nonumber\\
& &v^*\cdot \partial_c v - v\cdot \partial_{c}v^*
=0 \ , 
\nonumber\\
& &\gamma^*\cdot \partial_{\eta} \gamma - \gamma\cdot \partial_{\eta}\gamma^*
=0 \ , 
\nonumber\\
& &v^*\cdot \partial_{\eta} v - v\cdot \partial_{\eta}v^*
=\eta^*\cdot G_{cr} \ . 
\end{eqnarray}
With the aid of the relation (\ref{2-14}), the conditions (\ref{2-10}) 
and (\ref{2-11}), together with the relation (\ref{2-9}), lead to 
\begin{eqnarray}
& &F_{cr}=|\gamma|^2\left[
u^2(\Gamma'/\Gamma)_{ch}+|v|^2\left(
(\Gamma'/\Gamma)_{sh}-|\gamma|^{-2}\right)\right]^{-1} \ , 
\label{2-15}\\
& &G_{cr}=1 \ . 
\label{2-16}
\end{eqnarray}
The form (\ref{2-12}) with the relation (\ref{2-15}) gives us 
\begin{equation}\label{2-17}
2|c|^2=|\gamma|^2\left[
u^2(\Gamma'/\Gamma)_{ch}+|v|^2\left((\Gamma'/\Gamma)_{sh}-|\gamma|^{-2}
\right)\right] \ . 
\end{equation}
The relation (\ref{2-13}) with the form (\ref{2-16}) leads to 
\begin{equation}\label{2-18}
v=\eta \ .
\end{equation}
The right-hand side of the relation (\ref{2-17}) is a 
function of $|\gamma|^2$ and $|\eta|^2$ ($=|v|^2$), and then, by 
solving the relation (\ref{2-17}) inversely, we can determine 
$|\gamma|^2$ as a function of $|c|^2$ and $|\eta|^2$. 
Thus, by substituting $|\gamma|^2$ into the relation (\ref{2-15}), 
$F_{cr}$ can be determined as a function of $|c|^2$ and $|\eta|^2$ 
and with the use of the form (\ref{2-12}), 
$\gamma$ can be expressed as a function of $(c, c^*)$ and 
($\eta, \eta^*)$.

Next, let us investigate the relations to the $su(1,1)$-algebra. 
In (I), we discussed the expectation values of the set 
$({\hat \tau}_{\pm,0})$ defined as 
\begin{equation}\label{2-19}
{\hat \tau}_+=({\hat c}^*)^2/2 \ , \qquad
{\hat \tau}_-=({\hat c})^2/2 \ , \qquad
{\hat \tau}_0={\hat c}^*{\hat c}/2+1/4 \ . 
\end{equation}
The set $({\hat \tau}_{\pm,0})$ obeys the $su(1,1)$-algebra. 
%
%
It is instructive to investigate the expectation value of 
$({\hat \tau}_{\pm,0})$ for the $su(1,1)$-coherent state. 
The above realization provides the value $-3/16$ for the 
Casimir operator ${\hat C}={\hat \tau}_0^2-({\hat \tau}_-{\hat \tau}_+
+{\hat \tau}_+{\hat \tau}_-)/2$, namely, $C=t(t-1)=-3/16$. 
Thus, two lowest weight states occur, that is, 
$\ket{0}_0$ and $\ket{0}_1$ for $t=1/4$ and $3/4$, respectively. 
Therefore, two type of the $su(1,1)$-coherent state can be constructed 
on the two lowest weight states. 
The $su(1,1)$-coherent state based on $\ket{0}_0$ is defined by 
\begin{eqnarray}\label{P1}
& &\ket{\gamma}_0=(1-\gamma^*\gamma)^{1/4}
\exp(\gamma{\hat \tau}_+)\ket{0}_0 \ , \\
& &{\hat \tau}_-\ket{0}_0=0 \ , \qquad 
{\hat \tau}_0\ket{0}_0=\frac{1}{4}\ket{0}_0 \ . \nonumber
\end{eqnarray}
Here, this coherent state gives the expectation values as 
\begin{eqnarray}\label{P2}
& &(\tau_+)_0={}_0\bra{\gamma}{\hat \tau}_+\ket{\gamma}_0
=\frac{1}{2}\frac{\gamma^*}{1-\gamma^*\gamma}
=\sqrt{1/2+|c_0|^2}\ c_0^*\ , \nonumber\\
& &(\tau_-)_0={}_0\bra{\gamma}{\hat \tau}_-\ket{\gamma}_0
=\frac{1}{2}\frac{\gamma}{1-\gamma^*\gamma}
=c_0\sqrt{1/2+|c_0|^2}\ , 
\nonumber\\
& &(\tau_0)_0={}_0\bra{\gamma}{\hat \tau}_0\ket{\gamma}_0
=\frac{1}{2}\frac{\gamma^*\gamma}{1-\gamma^*\gamma}+\frac{1}{4}
=|c_0|^2+\frac{1}{4}\ ,
\end{eqnarray}
where we define $c_0=\sqrt{1/2}\cdot\gamma/\sqrt{1-\gamma^*\gamma}$. 
On the other hand, the $su(1,1)$-coherent state based on $\ket{0}_1$ 
is defined by 
\begin{eqnarray}\label{P3}
& &\ket{\gamma}_1=(1-\gamma^*\gamma)^{3/4}
\exp(\gamma{\hat \tau}_+)\ket{0}_1 \ , \\
& &{\hat \tau}_-\ket{0}_1=0 \ , \qquad 
{\hat \tau}_0\ket{0}_1=\frac{3}{4}\ket{0}_1 \ . \nonumber
\end{eqnarray}
Here, this coherent state gives the expectation values as 
\begin{eqnarray}\label{P4}
& &(\tau_+)_1={}_1\bra{\gamma}{\hat \tau}_+\ket{\gamma}_1
=\frac{3}{2}\frac{\gamma^*}{1-\gamma^*\gamma}
=\sqrt{3/2+|c_1|^2}\ c_1^* \ , \nonumber\\
& &(\tau_-)_1={}_1\bra{\gamma}{\hat \tau}_-\ket{\gamma}_1
=\frac{3}{2}\frac{\gamma}{1-\gamma^*\gamma}
=c_1\sqrt{3/2+|c_1|^2}\ , 
\nonumber\\
& &(\tau_0)_1={}_1\bra{\gamma}{\hat \tau}_0\ket{\gamma}_1
=\frac{3}{2}\frac{\gamma^*\gamma}{1-\gamma^*\gamma}+\frac{3}{4}
=|c_1|^2 +\frac{3}{4} \ ,
\end{eqnarray}
where we define $c_1=\sqrt{3/2}\cdot\gamma/\sqrt{1-\gamma^*\gamma}$. 
Here, $c_i$ is the dynamical canonical variable. 
In general, the expectation value 
of ${\hat \tau}_0$ has the form 
$(\tau_0)=|c|^2+t$. Therefore, $\ket{\gamma}_0$ 
and $\ket{\gamma}_1$, which give $C=-3/16$, provide 
the value $t=1/4$ and $3/4$, respectively. 
%
%
The unified form of $\ket{\gamma}_0$ and $\ket{\gamma}_1$ can be 
given similar to the state (\ref{2-6}) as 
\begin{equation}\label{P5}
\ket{\Psi}=\alpha\ket{\gamma}_0+\beta\ket{\gamma}_1 \ .
\end{equation}
Here, it should be noted that $\alpha$ can be taken as a real number and 
it can be expressed in terms of $\beta$ because of the 
normalization condition : 
\begin{equation}\label{P6}
\bra{\Psi}\Psi \rangle=\alpha^2+|\beta|^2=1 \ .
\end{equation}
Thus, $(\gamma, \gamma^*)$ and $(\beta, \beta^*)$ can be regarded as 
dynamical variables. 
The expectation values of $su(1,1)$-generators are obtained as 
\begin{eqnarray}\label{P7}
& &\bra{\Psi}{\hat \tau}_+\ket{\Psi}
=\frac{1}{2}\frac{\gamma^*}{1-\gamma^*\gamma}(\alpha^*\alpha+
3\beta^*\beta) \ , \nonumber\\
& &\bra{\Psi}{\hat \tau}_-\ket{\Psi}
=\frac{1}{2}\frac{\gamma}{1-\gamma^*\gamma}(\alpha^*\alpha+
3\beta^*\beta) \ , \nonumber\\
& &\bra{\Psi}{\hat \tau}_0\ket{\Psi}
=\frac{1}{4}\frac{1+\gamma^*\gamma}{1-\gamma^*\gamma}
(\alpha^*\alpha+
3\beta^*\beta) \ .
\end{eqnarray}
Further, we obtain the following relation : 
\begin{eqnarray}\label{P8}
& &\frac{1}{2}(\bra{\Psi}{\dot \Psi}\rangle - \bra{{\dot \Psi}}\Psi\rangle)
=
\frac{1}{2}(\beta^*{\dot \beta}-{\dot \beta}^*\beta) 
+\frac{1}{2}\cdot\frac{1}{2}\frac{\gamma^*{\dot \gamma}-{\dot \gamma}^*\gamma}
{1-\gamma^*\gamma}\cdot(\alpha^*\alpha+3\beta^*\beta) \ , \qquad
\end{eqnarray}
where $\alpha^*\alpha=1-|\beta|^2$ in Eqs.(\ref{P7}) and (\ref{P8}). 
Thus, we can introduce the set of canonical variables $(b,b^*)$ and 
$(c_{\Psi}, c_{\Psi}^*)$ as follows : 
\begin{eqnarray}\label{P9}
& &b=\beta \ , \qquad b^*=\beta^* \ , \nonumber\\
& &c_{\Psi}=\frac{1}{\sqrt{2}}\frac{\gamma}{\sqrt{1-\gamma^*\gamma}}
\sqrt{\alpha^*\alpha+3\beta^*\beta} \ , \quad
c_{\Psi}^*=\frac{1}{\sqrt{2}}\frac{\gamma^*}{\sqrt{1-\gamma^*\gamma}}
\sqrt{\alpha^*\alpha+3\beta^*\beta} \ . \qquad
\end{eqnarray}
By using these canonical variables, the expectation values in Eq.(\ref{P7}) 
can be expressed as 
\begin{eqnarray}\label{P10}
& &\bra{\Psi}{\hat \tau}_+\ket{\Psi}
=c_{\Psi}^*\sqrt{1/2+|b|^2+|c_{\Psi}|^2} \ , \nonumber\\
& &\bra{\Psi}{\hat \tau}_-\ket{\Psi}
=c_{\Psi}\sqrt{1/2+|b|^2+|c_{\Psi}|^2} \ , \nonumber\\
& &\bra{\Psi}{\hat \tau}_0\ket{\Psi}
=|c_{\Psi}|^2+|b|^2/2+1/4 \ .
\end{eqnarray}
Here, from the normalization condition (\ref{P6}), 
we obtain $0 \leq |b|\ (=|\beta|) \leq 1$. 
Thus, we conclude that $t$ has the values from $1/4$ to $3/4$.

According to the previous paper (I), that is Eq.(4.15) in (I), 
the boson coherent states with even and odd boson numbers respectively 
which we introduced, namely $\ket{ch}$ and $\ket{sh}$, provide the expectation 
value of ${\hat \tau}_0$ as 
\begin{eqnarray}\label{T1}
& &(\tau_0)_{ch}=\bra{ch}{\hat \tau}_0\ket{ch}
=\frac{1}{4}+\frac{|\gamma|^2}{2}\cdot\left(\frac{\Gamma'}{\Gamma}\right)_{ch}
=\frac{1}{4}+|c|^2 \ , \nonumber\\
& &(\tau_0)_{sh}=\bra{sh}{\hat \tau}_0\ket{sh}
=\frac{1}{4}+\frac{|\gamma|^2}{2}\cdot\left(\frac{\Gamma'}{\Gamma}\right)_{sh}
=\frac{3}{4}+|c|^2 \ . 
\end{eqnarray}
On the other hand, 
the expectation value of $({\hat \tau}_{\pm,0})$ for the state 
$\ket{cr}$, which we denote $(\tau_{\pm,0})_{cr}$, are 
simply expressed as 
\begin{equation}\label{2-20}
(\tau_{\pm,0})_{cr}=u^2(\tau_{\pm,0})_{ch} +|v|^2 (\tau_{\pm,0})_{sh} \ .
\end{equation}
The explicit forms of $(\tau_{\pm,0})_{ch}$ and $(\tau_{\pm,0})_{sh}$ 
are given in the form (4.10) and (4.11) in (I). 
In the present case, the relation (\ref{2-20}) with the relations 
(4.10), (4.11) in (I) and (\ref{2-17}) presents us 
\begin{equation}\label{2-21}
(\tau_0)_{cr}=|c|^2+|\eta|^2/2 +1/4 \ . 
\end{equation}
Since $|\eta|\ (=|v|)$ has the value between 0 and 1, 
the state $\ket{cr}$ thus gives the value from $t=1/4$ to 3/4, 
where $t=1/4+|\eta|^2/2$.

The interest in (II) is concerned with the expectation value 
of ${\hat c}$ itself. 
It is given in the following form : 
\begin{eqnarray}\label{2-22}
(c)_{cr}&=&
uv|\gamma|\cdot \sqrt{\Gamma_{ch}/\Gamma_{sh}}
\left({\tilde f}(N)\right)_{ch}
+uv^*(\gamma^2/|\gamma|)\cdot \sqrt{\Gamma_{sh}/\Gamma_{ch}}
\left({\tilde f}(N)\right)_{sh} \ .\qquad
\end{eqnarray}
The explicit form will be shown in some concrete examples.

\section{Two examples}

In this section, we show two concrete examples. One is the case 
discussed in \S 6 of (I) and the other in \S 7 of (I). 
Let us start in the following case treated in \S 6 of (I)~ : 
\begin{equation}\label{3-1}
{\tilde f}(2n)=2n+1 \ , \qquad
{\tilde f}(2n-1)=1 \ .
\end{equation}
In the treatment in (I), this case satisfies the relation of the 
Poisson bracket of the $su(1,1)$-algebra. 
The quantities $\Gamma_{ch}$, $\Gamma'_{ch}$, $\Gamma_{sh}$ 
and $\Gamma'_{sh}$ are calculated as 
\begin{subequations}\label{3-2}
\begin{eqnarray}
& &\Gamma_{ch}=(1-|\gamma|^4)^{-1/2} \ , \qquad
\Gamma'_{ch}=|\gamma|^2(1-|\gamma|^4)^{-3/2} \ , 
\label{3-2a}\\
& &\Gamma_{sh}=|\gamma|^2(1-|\gamma|^4)^{-3/2} \ , \quad
\Gamma'_{sh}=(1-|\gamma|^4)^{-3/2}+3|\gamma|^4(1-|\gamma|^4)^{-5/2} \ . 
\qquad
\label{3-2b}
\end{eqnarray}
\end{subequations}
Thus, we have 
\begin{subequations}\label{3-3}
\begin{eqnarray}
& &(\Gamma'/\Gamma)_{ch}=|\gamma|^2/(1-|\gamma|^4) \ , 
\label{3-3a}\\
& &(\Gamma'/\Gamma)_{sh}-|\gamma|^{-2}=3|\gamma|^2/(1-|\gamma|^4) \ . 
\label{3-3b}
\end{eqnarray}
\end{subequations}
With the use of the relation (\ref{3-3}), $F_{cr}$ defined in the form 
(\ref{2-15}) is given by 
\begin{eqnarray}\label{3-4}
F_{cr}&=&
|\gamma|^2\left[ u^2\cdot |\gamma|^2/(1-|\gamma|^4) + |v|^2\cdot 
3|\gamma|^2/(1-|\gamma|^4)\right]^{-1}
\nonumber\\
&=&\frac{1-|\gamma|^4}{1+2|v|^2} \ .
\end{eqnarray}
Then, the relation for expressing $|\gamma|^4$ as a function of 
$|c|^2$ and $|\eta|^2$ is obtained : 
\begin{equation}\label{3-5}
|\gamma|^4=2|c|^2\cdot \frac{1-|\gamma|^4}{1+2|\eta|^2} \ .
\end{equation}
The solution of Eq.(\ref{3-5}) and $F_{cr}$ are obtained as 
\begin{eqnarray}
& &|\gamma|^4=\frac{2|c|^2}{1+2|\eta|^2+2|c|^2} \ , 
\label{3-6}\\
& &F_{cr}=\frac{1}{1+2|\eta|^2+2|c|^2} \ .
\label{3-7}
\end{eqnarray}
Thus, $\gamma$ can be expressed in terms of $(c, c^*)$ 
as follows : 
\begin{equation}\label{3-8}
\gamma=\sqrt{c}\cdot \left[\sqrt[4]{1/2+|\eta|^2+|c|^2}
\right]^{-1} \ . 
\end{equation}

The relation (\ref{2-20}), with the aid of the relations 
(4.10) and (6.10) in (I), gives us 
\begin{equation}\label{3-9}
(\tau_-)_{cr}=(\gamma^2/2)\cdot\left(
u^2\cdot 1/(1-|\gamma|^4)+|v|^2\cdot 3/(1-|\gamma|^4)\right) \ .
\end{equation}
Substituting the relation (\ref{3-8}) into (\ref{3-9}), 
we have 
\begin{equation}\label{3-10}
(\tau_-)_{cr}=c\sqrt{1/2+|\eta|^2+|c|^2} \ .
\end{equation}
It may be interesting to see that $(\tau_{\pm,0})_{cr}$ shown 
in the relations (\ref{2-21}) and (\ref{3-10}) still 
satisfies the relation of the Poisson bracket 
of the $su(1,1)$-algebra with $t=1/4+|\eta|^2/2$ and 
$t$ takes the value between $1/4$ and $3/4$. 
This means that our present system is just mixture of the states 
with $t=1/4$ and $3/4$ and $|\eta|^2$ plays a role of 
the parameter determining the degree of the mixture. 
Thus, it is possible to regard $\ket{cr}$ as a unified form of 
$\ket{ch}$ and $\ket{sh}$. 
The above form was also discussed in Ref.\citen{two}.

Our interest is concerned with the expectation value 
$(c)_{cr}$ shown in the relation (\ref{2-22}). 
In the present case, $({\tilde f}(N))_{ch}$ and 
$({\tilde f}(N))_{sh}$ are given as 
\begin{subequations}\label{3-11}
\begin{eqnarray}
& &\left({\tilde f}(N)\right)_{ch}=(1-|\gamma|^4)^{-3/2} \ , 
\label{3-11a}\\
& &\left({\tilde f}(N)\right)_{sh}
=|\gamma|^2(1-|\gamma|^4)^{-3/2} \ . 
\label{3-11b}
\end{eqnarray}
\end{subequations}
Then, $(c)_{cr}$ can be expressed in the form 
\begin{equation}\label{3-12}
(c)_{cr}=u\left(\sqrt{1-|\gamma|^4}\right)^{-1}
(v^*\gamma^2+v) \ .
\end{equation}
With the use of the relation (\ref{3-8}), $(c)_{cr}$ is 
given as 
\begin{eqnarray}\label{3-13}
(c)_{cr}&=&\sqrt{1-|\eta|^2}\cdot \left(\sqrt{1/2+|\eta|^2}\right)^{-1}
\nonumber\\
& &\times\left(\eta\sqrt{1/2+|\eta|^2+|c|^2}+\eta^* c\right) \ .
\end{eqnarray}
This form is identical to that derived 
in Ref. \citen{two}.

Next, we discuss the case which corresponds to that in \S 7 in (I) : 
\begin{equation}\label{3-14}
{\tilde f}(2n)=\sqrt{2n+1} \ , \qquad
{\tilde f}(2n-1)=1 \ . 
\end{equation}
The quantities $\Gamma_{ch}$, $\Gamma'_{ch}$, 
$\Gamma_{sh}$ and $\Gamma'_{sh}$ are calculated as 
\begin{subequations}\label{3-15}
\begin{eqnarray}
& &\Gamma_{ch}=\exp (|\gamma|^4/2) \ , \qquad
\Gamma'_{ch}=|\gamma|^2\exp(|\gamma|^4/2) \ , 
\label{3-15a}\\
& &\Gamma_{sh}=|\gamma|^2\exp (|\gamma|^4/2) \ , \qquad
\Gamma'_{sh}=\exp(|\gamma|^4/2)+|\gamma|^4\exp(|\gamma|^4/2) 
\ . 
\label{3-15b}
\end{eqnarray}
\end{subequations}
Thus, $(\Gamma'/\Gamma)_{ch}$ and 
$(\Gamma'/\Gamma)_{sh}-|\gamma|^{-2}$ are given as 
\begin{equation}\label{3-16}
\left(\Gamma'/\Gamma\right)_{ch}=
\left(\Gamma'/\Gamma\right)_{sh}-|\gamma|^{-2}=|\gamma|^2 \ .
\end{equation}
Therefore, $F_{cr}$ is simply obtained in the form 
\begin{equation}\label{3-17}
F_{cr}=|\gamma|^2\left[
u^2\cdot |\gamma|^2+|v|^2\cdot |\gamma|^2\right]^{-1}=1 \ .
\end{equation}
Then, $|\gamma|^4$ and $\gamma$ are given in the form 
\begin{eqnarray}
& &|\gamma|^4=2|c|^2 \ , 
\label{3-18}\\
& &\gamma=\sqrt{\sqrt{2} c} \ . 
\label{3-19}
\end{eqnarray}
The relation (\ref{2-20}), together with the relation 
(7.9) in (I), gives us 
\begin{equation}\label{3-20}
(\tau_-)_{cr} \sim (1-|\eta|^2)c\sqrt{1/2+|c|^2} 
+|\eta|^2 c \sqrt{3/2+|c|^2} \ .
\end{equation}
The present case does not satisfy the relation of 
the Poisson bracket.

Our next problem is to calculate the expectation value $(c)_{cr}$. 
In this case, the following formula is useful : 
\begin{subequations}\label{3-21}
\begin{eqnarray}
\left({\tilde f}(N)\right)_{ch}&=& 
\sum_{n=0}^{\infty} \frac{\left(\frac{|\gamma|^4}{2}\right)^n}
{n!}\sqrt{2n+1}\ e^{-|\gamma|^4/2} \nonumber\\
&\sim&\sqrt{|\gamma|^4+1} \ , 
\label{3-21a}\\
\left({\tilde f}(N)\right)_{sh}&=& 
\sum_{n=0}^{\infty} 
\frac{\left(\frac{|\gamma|^4}{2}\right)^n}
{n!}\cdot e^{-|\gamma|^4/2} 
=1 \ .
\label{3-21b}
\end{eqnarray}
\end{subequations}
Then, $(c)_{cr}$ can be expressed in the form 
\begin{equation}\label{3-22}
(c)_{cr}=u(v^*\gamma^2 + v\sqrt{|\gamma|^4+1}) \ .
\end{equation}
The relation (\ref{3-12}) gives us 
\begin{equation}\label{3-23}
(c)_{cr}=\sqrt{2}\sqrt{1-|\eta|^2} \ (\eta\sqrt{1/2+|c|^2}
+\eta^* c) \ .
\end{equation}

\section{An extension from the conventional boson coherent state}

In this section, we treat the case 
\begin{equation}\label{4-1}
{\tilde f}(2n)=1 \ , \qquad
{\tilde f}(2n-1)=1\ . 
\end{equation}
The above corresponds to the case discussed in \S 3 in (I). 
It is a straightforward extension from the conventional 
boson coherent state under the idea of introducing 
the parameter $(v, v^*)$. 
In this case, $\Gamma_{ch}$, $\Gamma'_{ch}$, 
$\Gamma_{sh}$ and $\Gamma'_{sh}$ are calculated 
in the form
\begin{subequations}\label{4-2}
\begin{eqnarray}
& &\Gamma_{ch}=\cosh |\gamma|^2 \ , \qquad
\Gamma'_{ch}=\sinh |\gamma|^2 \ , 
\label{4-2a}\\
& &\Gamma_{sh}=\sinh |\gamma|^2 \ , \qquad
\Gamma'_{sh}=\cosh |\gamma|^2 \ . 
\label{4-2b}
\end{eqnarray}
\end{subequations}
The above relation gives us 
\begin{subequations}\label{4-3}
\begin{eqnarray}
& &\left(\Gamma'/\Gamma\right)_{ch}=\tanh |\gamma|^2 \ , 
\label{4-3a}\\
& &\left(\Gamma'/\Gamma\right)_{sh}
-|\gamma|^{-2}=|\gamma|^{-2}(
|\gamma|^2\coth |\gamma|^2-1) \ . 
\label{4-3b}
\end{eqnarray}
\end{subequations}
With the use of the relation (\ref{4-3}), 
$F_{cr}$ is obtained in the form 
\begin{equation}\label{4-4}
F_{cr}=\left[u^2\frac{\tanh |\gamma|^2}{|\gamma|^2}
+|v|^2\frac{|\gamma|^2\coth |\gamma|^2 -1}{|\gamma|^4}
\right]^{-1} \ .
\end{equation}
The parameter $|\gamma|^4$ can be determined through the relation 
$$
|\gamma|^4=2|c|^2\cdot F_{cr} \ , 
$$
i.e., 
\begin{equation}\label{4-5}
|\gamma|^4=2|c|^2\cdot
\left[u^2\frac{\tanh |\gamma|^2}{|\gamma|^2}
+|v|^2\frac{|\gamma|^2\coth |\gamma|^2 -1}{|\gamma|^4}
\right]^{-1} \ .
\end{equation}
However, we cannot determine $|\gamma|^4$ in terms of 
$|c|^2$. 
Then, in the same manner as that adopted in (I), 
we make approximation. 
Later, the idea will be shown. 

Since, in the present case, 
$({\tilde f}(N){\tilde f}(N+1))_{ch}=
({\tilde f}(N){\tilde f}(N+1))_{sh}=1$, the relation 
(4.10) in (I) gives us  
\begin{equation}\label{4-6}
(\tau_-)_{cr}=\gamma^2/2=c\sqrt{F_{cr}/2} \ . 
\end{equation}
In order to calculate $(c)_{cr}$, 
$\Gamma_{ch}/\Gamma_{sh}$ ($=\coth |\gamma|^2$) 
and $\Gamma_{sh}/\Gamma_{ch}$ ($=\tanh |\gamma|^2$) 
are necessary. 
With the use of the relation (\ref{4-5}), $|\gamma|^2$ 
can be expressed in terms of $|c|^2$ and $|\eta|^2$. 
Since, in this case, 
$({\tilde f}(N))_{ch}=({\tilde f}(N))_{sh}=1$, the form 
(\ref{2-22}) leads to 
\begin{equation}\label{4-7}
(c)_{cr}=\sqrt{1-|\eta|^2}
\left[\left(\sqrt{\frac{\tanh |\gamma|^2}{|\gamma|^2}}
\right)^{-1}\cdot \eta +
\sqrt{2F_{cr}}\sqrt{\frac{\tanh |\gamma|^2}{|\gamma|^2}}\cdot
\eta^* c\right] \ .
\end{equation}
Of course, $\sqrt{\tanh |\gamma|^2/|\gamma|^2}$ can 
be expressed as 
\begin{equation}\label{4-8}
\sqrt{\frac{\tanh |\gamma|^2}{|\gamma|^2}}=
\sqrt{\frac{\tanh (\sqrt{2}|c|\sqrt{F_{cr}})}
{\sqrt{2}|c|\sqrt{F_{cr}}}} \ .
\end{equation}
It may be clear from the relations (\ref{4-6}) and (\ref{4-7}) 
that $F_{cr}$ expressed in terms of $|c|^2$ and $|\eta|^2$ is 
necessary. This is our next task.

First, let us show an idea which is a straightforward extension of 
that given in (I). 
In the region $|\gamma|^2 \sim 0$, $F_{cr}$ can be 
approximated as 
\begin{eqnarray}\label{4-9}
F_{cr}&\sim&
(1-2|\eta|^2/3)^{-1}
+(1/3)(1-14|\eta|^2/15)(1-2|\eta|^2/3)^{-2}\cdot |\gamma|^4 
\nonumber\\
& &-(1/45)(1-4|\eta|^2/7-44|\eta|^4/105)
(1-2|\eta|^2/3)^{-3}\cdot |\gamma|^8 \ .
\end{eqnarray}
The above comes from the relation (\ref{4-4}). 
With the use of the relation (\ref{4-5}), 
iteratively, we obtain the following relation : 
\begin{eqnarray}\label{4-10}
|\gamma|^4&\sim&
(1-2|\eta|^2/3)^{-1}\cdot 2|c|^2
+(1/3)(1-14|\eta|^2/15)(1-2|\eta|^2/3)^{-3}
\cdot(2|c|^2)^2
\nonumber\\
& &+(4/45)(1-46|\eta|^2/21+376|\eta|^4/315)
(1-2|\eta|^2/3)^{-5}\cdot (2|c|^2)^3 \ ,
\end{eqnarray}
i.e., 
\begin{eqnarray}\label{4-11}
F_{cr}&\sim&
(1-2|\eta|^2/3)^{-1}
+(1/3)(1-14|\eta|^2/15)(1-2|\eta|^2/3)^{-3}
\cdot(2|c|^2)
\nonumber\\
& &+(4/45)(1-46|\eta|^2/21+376|\eta|^4/315)
(1-2|\eta|^2/3)^{-5}\cdot (2|c|^2)^2 \ .\quad
\end{eqnarray}
In the region $|\gamma|^2\rightarrow \infty$, the asymptotic 
form is given as 
\begin{equation}\label{4-12}
F_{cr}\longrightarrow 
|\gamma|^2 [u^2+|v|^2(1-|\gamma|^{-2})]
=|\gamma|^2(1-|\eta|^2/|\gamma|^2) \ .
\end{equation}
With the use of the relation (\ref{4-5}), we have 
\begin{equation}\label{4-13}
|\gamma|^2\longrightarrow 2|c|^2+|\eta|^2 \ , 
\end{equation}
i.e., 
\begin{equation}\label{4-14}
F_{cr}\longrightarrow 2|c|^2+2|\eta|^2 \ .
\end{equation}
Therefore, our problem is to find $F_{cr}$ which 
leads us to the form (\ref{4-11}) in the region 
$|c|^2\sim 0$ and to the form (\ref{4-14}) at the limit 
$|c|^2\rightarrow \infty$. 
In Appendix of (I), we showed a possible method. 
In this paper, we adopt the following form : 
\begin{eqnarray}\label{4-15}
F_{cr}&\sim& p \exp [-q\cdot (2|c|^2)-r\cdot (2|c|^2)^2] 
+2|c|^2+2|\eta|^2 \ .
\end{eqnarray}
If $q>0$ and $r>0$, the first term in the right-hand side 
of the form (\ref{4-15}) asymptotically vanishes and 
the second term remains. 
We can prove that, if $p$, $q$ and $r$ take the values shown 
in the following, $F_{cr}$ in the relation (\ref{4-15}) 
is reduced to the form (\ref{4-11}) : 
\begin{eqnarray}\label{4-16}
p&=&(1-2|\eta|^2/3)-(4/3)|\eta|^2(1-|\eta|^2) \ , 
\nonumber\\
q&=&(2/3)\left[
(1-44|\eta|^2/45)-(14/9)|\eta|^2(1-|\eta|^2)(1-2|\eta|^2/7)
\right]
\nonumber\\
& &\times (1-2|\eta|^2/3)^{-2}\left[
(1-2|\eta|^2/3)^{-2}-(4/3)|\eta|^2(1-|\eta|^2)\right]^{-1} \ , 
\nonumber\\
r&=&(2/15)\biggl[
(5/3)\left[(1-44|\eta|^2/45)-(14/9)|\eta|^2(1-|\eta|^2)
(1-2|\eta|^2/7)\right]^2 \nonumber\\
& &\times (1-2|\eta|^2/3)^{-4}\left[
(1-2|\eta|^2/3)-(4/3)|\eta|^2(1-|\eta|^2)\right]^{-2} 
\nonumber\\
& &-
(2/3)\left[
(1-314|\eta|^2/315)-(376/315)|\eta|^2(1-|\eta|^2)\right]
\nonumber\\
& &\times (1-2|\eta|^2/3)^{-4}\left[
(1-2|\eta|^2/3)-(4/3)|\eta|^2(1-|\eta|^2)\right]^{-1} 
\biggl]
\ . 
\end{eqnarray}
Thus, we can express the expectation value $(\tau_-)_{cr}$ and 
$(c)_{cr}$ approximately in terms of 
$(c, c^*)$ and $(\eta, \eta^*)$.

\section{Deformation in the boson mapping method}

In Part (I), we formulated the boson-pair coherent states for the 
cases $u=1$ and $0$ independently and the relation 
to the MYT boson mapping method was discussed. 
In this section, we contact with the MYT boson mapping for the 
present case. 
For this problem, we have already developed this investigation 
in the case (\ref{3-1})\cite{two} and, in this sense, 
the form treated in this section is a natural generalization 
from that developed in Ref. \citen{two}.

Following Part (I) and Ref.\citen{two}, we introduce 
the following orthogonal set : 
\begin{equation}\label{5-1}
\rket{-n}=\rket{n}_d \ , \qquad \rket{+n}={\hat \zeta}\rket{n}_d \ .
\end{equation}
The state $\rket{n}_d$ denotes 
\begin{equation}\label{5-2}
\rket{n}_d=(\sqrt{n!})^{-1}({\hat d}^*)^n \rket{0}_d \ .\qquad
({\hat d}\rket{0}_d=0 )
\end{equation}
Here, $({\hat d} , {\hat d}^*)$ is a set of a boson operator. 
Further, the operator $({\hat \zeta} , {\hat \zeta}^*)$ 
denotes a set of a 
fermion operator obeying 
\begin{equation}\label{5-3}
\{\ {\hat \zeta} \ , \ {\hat \zeta}^*\ \}=1\ , \qquad
{\hat \zeta}^2={\hat \zeta}^{*2}=0 \ , 
\qquad
[\ {\hat \zeta} \ , \ {\hat d}\ ]=
[\ {\hat \zeta} \ , \ {\hat d}^*\ ]=0\ . 
\end{equation}
We set up the relation between the original and the new 
space in the following way : 
\begin{equation}\label{5-4}
\ket{2n} \sim \rket{-n} \ , 
\qquad
\ket{2n+1} \sim \rket{+n} \ . 
\end{equation}
Then, the mapping operator ${\hat U}$ characterizing 
the MYT boson mapping is introduced in the form 
\begin{equation}\label{5-5}
{\hat U}=\sum_{n=0}^{\infty} \left(
\rket{-n}\bra{2n} + \rket{+n}\bra{2n+1} \right) \ .
\end{equation}
Clearly, there exists the relation 
\begin{equation}\label{5-6}
{\hat U}{\hat U}^{\dagger} = {\hat U}^{\dagger}{\hat U}=1 \ .
\end{equation}
The operators $({\hat \tau}_{\pm,0})$ and ${\hat c}$ are mapped on the 
following form : 
\begin{equation}\label{5-7}
{\tilde \tau}_{\pm,0}={\hat U}{\hat \tau}_{\pm,0}{\hat U}^{\dagger} \ ,
\qquad
{\tilde c}={\hat U}{\hat c}{\hat U}^{\dagger} \ .
\end{equation}
Explicitly, they are given as 
\begin{eqnarray}
& &{\tilde \tau}_+ =
{\hat d}^*
\sqrt{1/2+{\hat \zeta}^*{\hat \zeta}+{\hat d}^*{\hat d}} \ , 
\nonumber\\
& &{\tilde \tau}_- =
\sqrt{1/2+{\hat \zeta}^*{\hat \zeta}+{\hat d}^*{\hat d}}\ 
{\hat d} \ , 
\nonumber\\
& &{\tilde \tau}_0 =
{\hat d}^*{\hat d}+1/4+{\hat \zeta}^*{\hat \zeta}/2 \ , 
\label{5-8}\\
& &{\tilde c}=\left(\sqrt{1/2+{\hat \zeta}^*{\hat \zeta}
+{\hat d}^*{\hat d}}\ {\hat d}+{\hat \zeta}^*{\hat d}\right)
\cdot\sqrt{(1-{\hat \zeta}^*{\hat \zeta})/
(1/2+{\hat \zeta}^*{\hat \zeta})} \ .
\label{5-9}
\end{eqnarray}
The expressions (\ref{5-8}) and (\ref{5-9}) have been derived 
in Ref.\citen{two}.

The operator ${\hat U}$ determines the image of $\ket{cr}$, 
which we denote as $\rket{cr}$ : 
\begin{equation}\label{5-10}
\rket{cr}=\left(\sqrt{\Gamma_{cr}}\right)^{-1} u 
\exp\left[
\frac{\gamma^2}{\sqrt{2}}\cdot{\hat d}^*\frac{
{\tilde f}(2{\hat M}+{\hat \mu}){\tilde f}(2{\hat M}+{\hat \mu}+1)}
{\sqrt{2{\hat M}+2{\hat \mu}+1}}\right]
\rket{m} \ , 
\end{equation}
\vspace{-0.2cm}
\setcounter{equation}{9}
\begin{subequations}\label{5-10a}
\begin{equation}
\rket{m}=\exp \left[\frac{|\gamma|v}{u}\sqrt{
\frac{\Gamma_{ch}}{\Gamma_{sh}}}\cdot {\hat \zeta}^*\right]
\rket{0} \ ,
\end{equation}
\end{subequations}
\vspace{-0.2cm}
\begin{equation}\label{5-11}
{\hat M}={\hat d}^*{\hat d} \ , \qquad
{\hat \mu}={\hat \zeta}^*{\hat \zeta} \ .
\end{equation}
For the case (\ref{3-1}), i.e., 
${\tilde f}(2n)=2n+1$ and 
${\tilde f}(2n-1)=1$, we have 
\begin{equation}\label{5-12}
\rket{cr}=\sqrt[4]{1-|\gamma|^4}\ u
\exp\left(\gamma^2\cdot{\hat d}^*
\sqrt{{\hat M}+1/2+{\hat \mu}}\right) \rket{m} \ , 
\end{equation}
\vspace{-0.2cm}
\setcounter{equation}{11}
\begin{subequations}\label{5-12a}
\begin{equation}
\rket{m}=\exp\left[\frac{v}{u}\sqrt{1-|\gamma|^4}
\cdot{\hat \zeta}^*\right]
\rket{0} \ . 
\end{equation}
\end{subequations}
The case (\ref{3-14}), i.e., 
${\tilde f}(2n)=\sqrt{2n+1}$ and 
${\tilde f}(2n-1)=1$, gives us 
\begin{equation}\label{5-13}
\rket{cr}=\exp(-|\gamma|^4/4)\ u
\exp\left(\gamma^2/\sqrt{2}\cdot{\hat d}^*\right)
\rket{m} \ , 
\end{equation}
\vspace{-0.2cm}
\setcounter{equation}{12}
\begin{subequations}\label{5-13a}
\begin{equation}
\rket{m}=\exp\left[\frac{v}{u}\cdot{\hat \zeta}^*\right]
\rket{0} \ . 
\end{equation}
\end{subequations}
Further, the case (\ref{4-1}), i.e., 
${\tilde f}(2n)=1$ and 
${\tilde f}(2n-1)=1$, leads us to 
\begin{equation}\label{5-14}
\rket{cr}=\left(\sqrt{\cosh |\gamma|^2}\right)^{-1} 
\ u
\exp\left[\gamma^2/{2}\cdot{\hat d}^*
\left(\sqrt{{\hat M}+1/2+{\hat \mu}}\right)^{-1}\right]
\rket{m} \ , 
\end{equation}
\vspace{-0.2cm}
\setcounter{equation}{13}
\begin{subequations}\label{5-14a}
\begin{equation}
\rket{m}=\exp\left[\frac{\gamma v}{u}
\sqrt{\coth |\gamma|^2}\cdot{\hat \zeta}^*\right]
\rket{0} \ . 
\end{equation}
\end{subequations}
The above three states show us that they are the quite natural generalization 
of the forms discussed in \S 8 in (I). 
The form (\ref{5-12}) is of the same form as that discussed 
in Ref. \citen{two}.

\section{Discussion}

In \S 3, we have shown two examples which are the exact inverses 
of the relation (\ref{2-17}), and in \S 4 the approximate relation 
has been shown. 
In this section, much simpler forms than the forms given in the previous 
sections are discussed, while the form given in \S 4 
is too complicated. 

For this purpose, we introduce two functions 
for $|\gamma|^2$, namely, $(\tanh |\gamma|^2)_1$ and 
$(\tanh |\gamma|^2)_2$, in the following forms : 
\begin{eqnarray}
& &(\tanh |\gamma|^2)_1 = |\gamma|^2/\sqrt{1+|\gamma|^4} \ ,
\label{6-1}\\
& &(\tanh |\gamma|^2)_2 = |\gamma|^2/
(\sqrt{9/4+|\gamma|^4}-1/2) \ .
\label{6-2}
\end{eqnarray}
The behaviors of $\tanh |\gamma|^2$, $(\tanh |\gamma|^2)_1$ 
and $(\tanh |\gamma|^2)_2$ in the regions $|\gamma|^2\sim 0$ and 
$|\gamma|^2 \rightarrow \infty$ are summarized as follows : \\
(1) In the region $|\gamma|^2 \sim 0$ : 
\begin{subequations}\label{6-3}
\begin{eqnarray}
& &\tanh |\gamma|^2 \sim |\gamma|^2 (1-(1/3)|\gamma|^4) \ , 
\label{6-3a}\\
& &(\tanh |\gamma|^2)_1 \sim |\gamma|^2 (1-(1/2)|\gamma|^4) \ , 
\label{6-3b}\\
& &(\tanh |\gamma|^2)_2 \sim |\gamma|^2 (1-(1/3)|\gamma|^4) \ . 
\label{6-3c}
\end{eqnarray}
\end{subequations}
(2) In the region $|\gamma|^2 \rightarrow \infty$ : 
\begin{subequations}\label{6-4}
\begin{eqnarray}
& &\tanh |\gamma|^2 \sim 1-\varepsilon \ , 
\label{6-4a}\\
& &(\tanh |\gamma|^2)_1 \sim 1-\varepsilon \ , 
\label{6-4b}\\
& &(\tanh |\gamma|^2)_2 \sim 1+\varepsilon \ . 
\label{6-4c}
\end{eqnarray}
\end{subequations}
Here, $\varepsilon$ denotes a positive infinitesimal 
parameter. 
Then, concerning $\tanh |\gamma|^2$, we can conclude that in 
the region $|\gamma|^2 \sim 0$, the behavior of 
$(\tanh |\gamma|^2)_2$ is nearer that of 
$(\tanh |\gamma|^2)_1$ and in the region 
$|\gamma|^2 \rightarrow \infty$, the situation is in reversal. 
For $\coth |\gamma|^2$, we define $(\coth |\gamma|^2)_1$ 
and $(\coth |\gamma|^2)_2$ in the form 
\begin{eqnarray}
& &(\coth |\gamma|^2)_1=(\tanh |\gamma|^2)_1^{-1}
=\sqrt{1+|\gamma|^4}/|\gamma|^2 \ , 
\label{6-5}\\
& &(\coth |\gamma|^2)_2=(\tanh |\gamma|^2)_2^{-1}
=\left(\sqrt{9/4+|\gamma|^4}-1/2\right)/|\gamma|^2 \ . 
\label{6-6}
\end{eqnarray}

Under the above preparation, for two cases, 
we find $|\gamma|^4$ expressed in terms of $2|c|^2$. 
By substituting the forms (\ref{6-1}) and (\ref{6-5}) 
into the relation (\ref{4-5}), 
we have 
\begin{equation}\label{6-7}
2|c|^2=|\gamma|^4
\left(\frac{u^2}{\sqrt{1+|\gamma|^4}}
+\frac{|v|^2}{\sqrt{1+|\gamma|^4}+1}\right) \ .
\end{equation}
The relation (\ref{6-7}) can be inversely solved 
in exact form, and through the relation 
$|\gamma|^4=2|c|^2\cdot (F_{cr})_1$, we obtain 
\begin{eqnarray}\label{6-8}
(F_{cr})_1&=&(1/2)\biggl[
(2|c|^2+2|\eta|^2)\cdot
\left(1+\frac{|\eta|^2}{\sqrt{(2|c|^2+|\eta|^2)^2+
4(1-|\eta|^2)}+(2-|\eta|^2)}\right) 
\nonumber\\
& &\qquad\quad +\sqrt{(2|c|^2+|\eta|^2)^2+4(1-|\eta|^2)}\biggl] \ .
\end{eqnarray}
Further, substituting the forms (\ref{6-2}) and (\ref{6-6}) 
into the relation (\ref{4-5}), we have 
\begin{equation}\label{6-9}
2|c|^2=|\gamma|^4
\left(\frac{u^2}{\sqrt{9/4+|\gamma|^4}-1/2}
+\frac{|v|^2}{\sqrt{9/4+|\gamma|^4}+3/2}\right) \ .
\end{equation}
The relation (\ref{6-9}) gives us 
\begin{eqnarray}\label{6-10}
(F_{cr})_2&=&(1/2)\biggl[
1+(2|c|^2-2(1-2|\eta|^2))\nonumber\\
& &\qquad\quad
\times 
\left(1+\frac{2|\eta|^2}{\sqrt{(2|c|^2-(1-2|\eta|^2))^2+
8(1-|\eta|^2)}+(3-2|\eta|^2)}\right) 
\nonumber\\
& &\qquad\quad +\sqrt{(2|c|^2-(1-2|\eta|^2))^2+
8(1-|\eta|^2)}\biggl] \ .
\end{eqnarray}

In the regions $2|c|^2 \sim 0$ and $2|c|^2 \rightarrow \infty$, 
$F_{cr}$, $(F_{cr})_1$ and $(F_{cr})_2$ are approximated as follows : \\
(1) In the region $2|c|^2 \sim 0$ : 
\begin{subequations}\label{6-11}
\begin{eqnarray}
& &F_{cr}\sim (1-2|\eta|^2/3)^{-1}
+(1/3)(1-14|\eta|^2/15)(1-2|\eta|^2/3)^{-3}\cdot 2|c|^2 \ , 
\label{6-11a}\\
& &(F_{cr})_1\sim \frac{2}{1-|\eta|^2}
+\frac{4-3|\eta|^2}{(2-|\eta|^2)^3}\cdot 2|c|^2 \ , 
\label{6-11b}\\
& &(F_{cr})_2 \sim (1-2|\eta|^2/3)^{-1}
+(1/3)(1-8|\eta|^2/9)(1-2|\eta|^2/3)^{-3}\cdot 2|c|^2 \ . \quad
\label{6-11c}
\end{eqnarray}
\end{subequations}
(2)In the region $2|c|^2\rightarrow \infty$ : 
\begin{subequations}\label{6-12}
\begin{eqnarray}
& &F_{cr} \longrightarrow 2|c|^2+ 2|\eta|^2 \ , 
\label{6-12a}\\
& &(F_{cr})_1\longrightarrow 2|c|^2+2|\eta|^2 \ , 
\label{6-12b}\\
& &(F_{cr})_2 \longrightarrow 2|c|^2-1+3|\eta|^2 \ . 
\label{6-12c}
\end{eqnarray}
\end{subequations}
It may be interesting to see that the relations (\ref{6-11}) 
and (\ref{6-12}) are in the situation similar to the 
relations (\ref{6-3}) and (\ref{6-4}). 
However, it should be noted that 
the form (\ref{6-8}) or (\ref{6-10}) are simple, and 
then, as a possible approximation, it may be useful 
for the case of the state $\ket{cr}$ characterized by the 
relation (\ref{4-1}).

Finally, we give a short comment on the functions 
$(\tanh |\gamma|^2)_i$ and $(\coth |\gamma|^2)_i$ 
$(i=1,2)$. 
These were introduced as numerical approximation of 
$\tanh |\gamma|^2$ and $\coth |\gamma|^2$. 
Then, forgetting the above-mentioned background, 
it may be interesting to investigate a problem 
which $\ket{cr}$ produces $(\tanh |\gamma|^2)_i$ and 
$(\coth |\gamma|^2)_i$ as $(\Gamma'/\Gamma)_{ch,i}$ 
and $(\Gamma'/\Gamma)_{sh,i}$ ($i=1,2$), respectively. 
By integrating $(\tanh |\gamma|^2)_i$ and 
$(\coth |\gamma|^2)_i$ by $|\gamma|^2$, 
we have 
\begin{subequations}\label{6-13}
\begin{eqnarray}
& &\Gamma_{ch,1}=\exp{\left[|\gamma|^4/(1+\sqrt{1+|\gamma|^4})\right]} \ , 
\nonumber\\
& &\Gamma_{sh,1}
=|\gamma|^2\cdot 2/(1+\sqrt{1+|\gamma|^4})
\exp{\left[|\gamma|^4/(1+\sqrt{1+|\gamma|^4})\right]} \ , 
\label{6-13a}\\
& &\Gamma_{ch,2}
=\sqrt{\sqrt{9/4+|\gamma|^4}-1/2} \ 
\exp{\left[|\gamma|^4/(3/2+\sqrt{9/4+|\gamma|^4})\right]} \ , 
\nonumber\\
& &\Gamma_{sh,2}
=|\gamma|^2\! \left(\!
\sqrt{3/(3/2+\sqrt{9/4+|\gamma|^4})}\!\right)^3\!\! 
\exp{\!\left[|\gamma|^4/(3/2+\sqrt{9/4+|\gamma|^4})\right]} \ . \qquad\quad
\label{6-13b}
\end{eqnarray}
\end{subequations}
As is clear from the relation (\ref{2-3}), 
$\Gamma_{ch,i}$ and $\Gamma_{sh,i}$ should be 
regarded as the forms obtained by the 
sum of the power series expansion for $|\gamma|^2$. 
However, such functions as $\sqrt{1+|\gamma|^4}$ and 
$\sqrt{9/4+|\gamma|^4}$ cannot be 
expanded for $|\gamma|^2$ in the region $|\gamma|^2 >1$. 
Therefore, in the present framework where $\dket{ch}$ and 
$\dket{sh}$ are of the forms shown in the relation 
(\ref{2-2}), our idea is powerless, and we have to 
give the negative answer inevitably. 
In Part (III), this problem will be discussed 
where ${\tilde f}(2n-1)$ and ${\tilde f}(2n)$ 
characterizing the deformation depend on the parameter 
$|\gamma|^2$.

\end{document}